\documentstyle{article}

\input eqalign.sty
\def\air{\rightarrow}
\def\p{$p$\--adic~}
\def\ze{|\xi |_p}

\begin{document}
\begin{flushright}
October 1991 \\
EFI 91-43 \\
\end{flushright}

\begin{center}
MACDONALD POLYNOMIALS FROM SKLYANIN ALGEBRAS: \\
A CONCEPTUAL BASIS FOR THE $p$-ADICS-QUANTUM GROUP \\
CONNECTION\footnote{Work supported in part by the NSF: PHY-9000386} \\

\end{center}
\bigskip
\medskip

\begin{center}
Peter G. O. Freund \\

\smallskip
{ \em Enrico Fermi Institute and Department of Physics\\
University of Chicago, Chicago, IL 60637 \\ }

\medskip

and \\

\medskip

Anton V.~Zabrodin \\

\smallskip
{\em Institute of Chemical Physics \\
Kosygina Str. 4, SU-117334, Moscow USSR \\ }
\end{center}

\baselineskip = 12pt
\centerline{ABSTRACT}

\begin{quote}

We establish a previously conjectured connection between $p$-adics and
quantum groups.
We find in Sklyanin's two parameter elliptic quantum algebra and its
generalizations,  the conceptual basis for the Macdonald polynomials,
which ``interpolate'' between the zonal spherical functions of related real
and $p$\--adic symmetric spaces.
The elliptic quantum algebras underlie the $Z_n$\--Baxter models.
We show that in the  $n \air \infty$ limit, the Jost function for the
scattering of {\em first} level excitations in the $Z_n$\--Baxter model
coincides with the Harish\--Chandra\--like $c$\--function constructed from
the Macdonald polynomials associated to the root system $A_1$.
The partition function of the $Z_2$\--Baxter model itself is also expressed
in terms of
this Macdonald\--Harish\--Chandra\ $c$\--function, albeit in a
less simple way.
We relate the two parameters $q$ and $t$ of the Macdonald polynomials to the
anisotropy and modular parameters of the Baxter model.
In particular the $p$\--adic ``regimes'' in the Macdonald polynomials
correspond to a discrete sequence of XXZ models.
We also discuss the possibility of ``$q$\--deforming'' Euler products.

\end{quote}

\baselineskip = 16pt
\section{Introduction}

A connection between $p$-adics and quantum deformations has been
noticed [1-5] in a
variety of contexts over the past few years.
The possibility of such a connection emerges from work on
\p strings [6,7] and
$q$\--strings [8,9];
from work on  scattering on real [10,11],
\p [4] and quantum [12]  symmetric spaces; and
from work on Macdonald polynomials associated to ``admissible pairs'' of root
systems [1,2].

All this evidence points in the direction of   quantum
group\--like objects
with {\em two} deformation parameters and
the corresponding quantum symmetric spaces as underlying this
``$p$\--adics
quantum deformation connection'' [2,3,12].
Essentially, this is
how such a connection expected to work.
Corresponding to a root system $R$ (or more generally to an ``admissible
pair'' of root systems), one constructs a two parameter family of quantum
symmetric spaces, such that their
zonal spherical functions (zsf's) ``interpolate'' between the zsf's of
ordinary real and \p symmetric spaces in the following sense [1,2].
If we call the two parameters $q$ and $t$ then

a) for $q=0, ~ t = 1/p, ~ p =$ prime, these zsf's essentially reduce
to the zsf's of a \p symmetric space whose restricted root system is
$R^\vee$, the dual of the chosen root system $R$;

b) for $t = q^l \air 1$, with a certain value of $l$, these zsf's reduce
to the zsf's of the real symmetric space with restricted root system $R$.

There has been some progress [12-15] in exploring property b) in the
context of
one\--parameter quantum groups, obtained form the full two\--parameter groups
by imposing $t = q^l$, though not necessarily with $t$ near one.
However, none of the \p cases of  a) above can be reached this way.
We therefore have to address the full two\--parameter problem.

Here we
shall do just that and find that the {\em two parameter quantum algebra of
Sklyanin and its generalizations provide the conceptual understanding of
the $p$\--adics\--quantum deformation connection}.  Specifically, we
shall consider the $Z_n$\--Baxter model of statistical mechanics [16-19]
on a square
lattice for which the underlying algebra is of the (generalized) Sklyanin
type [20-24].
For this model,
in a regime such that the equivalent magnetic chain is antiferromagnetic
with finite gap,
we will study the scattering of two {\em first} level
excitations and will find that, in the $n \air \infty$ limit, the
corresponding Jost function coincides with the Harish\--Chandra\--like
$c$\--function [25]  obtained from Macdonald's polynomials for root system
$A_1$
(see Eq.~(5.8)).
The anisotropy parameter and the modular parameter of the Baxter model, are
then related with the parameters $q$ and $t$ according to the relations
(5.9).
This way of establishing the connection is like ``fishing'' for $SU(2)$
inside $SU(\infty )$.
One can also establish a connection directly between the $Z_2$\--Baxter
model and the $A_1$\--Macdonald\--Harish\--Chandra $c$\--function, but this
relation is less simple (see Eqs.~(5.10)).
These connections are our main results.
We suspect that both, the complexity of Eqs.~(5.10), and the need for the
$n \air \infty$ limit before the transparent Eq.~(5.8) is captured, are
connected with the involved coproduct situation in elliptic quantum
algebras [24].
One may wonder what physics corresponds to $(q,t) = (0, 1/p)$ in which cases
the Macdonald polynomials yield zsf's of \p symmetric spaces (case a)
above).
It is readily checked that the choice of parameters $(q, t ) = ( 0, 1/p)$
corresponds to
an XXZ model, with a particular value of the anisotropy parameter.

Mathematically, the most remarkable feature of our result is that
Sklyanin's elliptic quantum group
and its generalizations,  unify the $p$-adic and real versions
of  a  Lie group ($SL(2)$ in this case).  Of course a unification of
$SL(2, {\bf R})$ with the $SL(2, {\bf Q}_p )$'s occurs also in the
adelic [26]
context.
But the
unification which we have in mind here is of a completely different
nature.  It does not involve Euler products, but rather two real
parameters which can be ``dialed" for any archimedean or non-archimedean
case.  One can nevertheless ask the question about how this new
unification relates to the adelic one.  We shall therefore discuss the
possibility of $q$-deforming Euler products!

\section{Macdonald polynomials for the root system $A_1$}

Starting from any ``admissible pair" $(R,S)$ of root systems [27],
Macdonald [1,2] has constructed a corresponding family of orthogonal
polynomials enjoying some truly remarkable features.  This construction
is very general, the root system [27] $R$ need not even be reduced, it can
be
of the $BC_n$ type.
For our purposes,  it will suffice  to describe the Macdonald
construction in the simplest of all cases, where both $R$ and $S$ are
reduced and of rank 1, so that $R = S = A_1$.

The root system $A_1$ has one positive root $\alpha$ and one negative root
$- \alpha$.
The root lattice is $\Lambda_r = \{ n \alpha: n \in {\bf Z} \}$	and its
positive ``side" is $\Lambda_r^+ = \{ n \alpha : n \in {\bf Z} , n > 0
\}$.
The weight lattice $\Lambda$ of $A_1$ is $\Lambda = \{ {n \over 2} \alpha : n
\in {\bf Z} \}$ and the set of dominant weights is
$\Lambda^+ = \{ {n \over 2} \alpha : n \in {\bf Z} , n \geq 0 \}$.

Obviously $\Lambda_r^+ \subset \Lambda^+$.
On the weight lattice $\Lambda$ a partial order is defined
$$
\lambda_n > \lambda_m \longleftrightarrow \lambda_n - \lambda_m
\in \Lambda_r^+
\eqno(2.1a)
$$
or more explicitly
$$
{n \over 2} \alpha > {m \over 2} \alpha \longleftrightarrow 0 < n-m
 \in 2 {\bf Z}
\eqno(2.1b)
$$
The Weyl group $W$ of $A_1$ is $W = Z_2 = \{ 1, \sigma \}$, with
$1$ the identity element and $\sigma$ the reflection which takes $\pm \alpha$
into $\mp \alpha$.
The weight lattice $\Lambda$ is an abelian group under addition.    Its
group algebra $A$ over ${\bf R}$ is suggestively presented in terms of formal
exponentials
$$
\lambda = {n \over 2} \alpha \in \Lambda \longrightarrow
e^\lambda = e^{{n \over 2} \alpha} \in A
\eqno(2.2)
$$
so that
$$
e^{{m \over 2} \alpha} e^{{n \over 2} \alpha} = e^{\frac{m+n}{2} \alpha}~~~
\left ( e^{{n \over 2} \alpha} \right )^{-1}
= e^{-{n \over 2} \alpha } ~~~e^0 = 1
\eqno(2.3)
$$

These $e^{{n \over 2} \alpha}$ form an ${\bf R}$ basis of $A$.
The Weyl group action on $\Lambda$ defines
a $W$-action also on the group algebra $A$
$$
w (e^\lambda ) : = e^{w \lambda}~~~~~~ w \in W,~
\lambda \in \Lambda , ~e^\lambda \in A
\eqno(2.4)
$$

The Weyl-invariant elements of $A$ span a subalgebra
$$
A^W = \{ a \in A : wa=a ~,~~~~\forall w \in W \}
\eqno(2.5)
$$
of $A$.
Obviously the elements
$$
m_n = e^{{n \over 2} \alpha} + e^{- {n \over 2} \alpha} ~~~~~~ n \in
{\bf Z}_+
\eqno(2.6)
$$
provide an ${\bf R}$-basis of $A^W$.

Define the Weyl characters
$$
\chi_n =
\frac{e^{(n+1){\alpha \over 2}} - e^{- (n+1){\alpha \over 2}}}{e^{\alpha
\over 2} - e^{- {\alpha \over 2}}}
\eqno(2.7)
$$
Then the
$$
\chi_n ~~ {\rm with} ~~~ n \geq 0
\eqno(2.8)
$$
also provide an ${\bf R}$-basis of $A^W$.

Beside these ${\bf R}$-bases of $A^W$, there exists a much less obvious
two\--parameter
family of ${\bf R}$ bases of $A^W$.
This family of Macdonald bases comes into being due to the existence of a two
parameter family of positive\--definite scalar products on $A$.
They are constructed as follows.
Call the two real parameters $t$ and $q$ and consider the element $\Delta
(t, q)$ in $A$ defined by
$$
\Delta (t,q) =
\frac{(e^\alpha; q)_\infty}{(te^\alpha; q)_\infty}
\frac{(e^{- \alpha}; q)_\infty}{(te^{-\alpha}; q)_\infty}
\eqno(2.9)
$$
where $e^{\pm \alpha}$ are the elements in $A$ corresponding to the roots
$\pm \alpha$.
Here and throughout this paper we adopted the notation [28, 29]
$$
(a; q)_n = (1-a) (1-aq) \ldots (1 - a q^{n-1})
\eqno(2.10a)
$$
and
$$
(a; q)_\infty = \prod_{k=0}^{\infty} (1 - a q^k ),
\eqno(2.10b)
$$
so that
$$
(a; q)_n = \frac{(a; q)_\infty}{(aq^n; q)_\infty} ~~.
\eqno(2.10c)
$$
To each
$$
f = \sum_{\lambda \in \Lambda} f_\lambda e^\lambda \in A ~~~~
(f_\lambda \in {\bf R})
\eqno(2.11a)
$$
we associate its ``conjugate''
$$
\bar{f} = \sum_{\lambda \in \Lambda} f_\lambda e^{- \lambda} \in A
{}~.
\eqno(2.11b)
$$
Now we consider the 1-torus (circle) $T = {\bf R}/\Lambda_r^\vee$, where
$\Lambda_r^\vee$ is the root lattice of the dual root system.
Obviously any $x \in {\bf R}$ then has an image $x_T$ on the circle
$T$.
Each $e^\lambda \in A$ can therefore be viewed as a character of $T$ via
$$
e^\lambda (x_T) = e^{i 2 \pi < \lambda , x >} ~~.
\eqno(2.12)
$$
Macdonald's two parameter family of positive-definite scalar products on $A$
is then given by
$$
< f , g>_{t,q} = {1 \over 2} \int_T f \bar{g} \Delta (t, q) ,
\eqno(2.13)
$$
the measure on $T$ being the (normalized) Haar measure.

Finally, for each scalar product $< ~, ~>_{t,q}$ in the family (2.13), we
define a {\em Macdonald basis} of $A^W$ by the following three requirements

a)
$$
P_m = m_m + \sum_{\stackrel{0 \leq n < m}{m-n \in 2 {\bf Z}}}
a_{m n} (q,t) m_n;
$$

b) all $a_{m n} (q,t)$ are rational functions of the two real parameters $q$
and $t$;

c) the $P_m$'s are orthogonal
under the Macdonald scalar product $ < , >_{t,q}$ on $A$, i.e.
$<P_m , P_n >_{t,q} = 0$ for $m \neq n$

These $P_n$ are clearly polynomials in
$m_1 = e^{\alpha /2} + e^{- \alpha / 2}$, or equivalently Laurent
polynomials in $e^{\alpha \over 2}$.
For the $A_1$ case under discussion, they are explicitly given in
terms of the famous Ro\-gers\--As\-key\--Is\-mail (RAI)
polynomials [30,31]
$C_n (x; t | q)$.
Specifically
$$
P_n \left ( e^{\alpha \over 2}; t | q \right ) =
\frac{(q; q)_n}{(t;q)_n} \Phi_n (e^{\alpha \over 2}; t | q )
\eqno(2.14a)
$$
where
$$
\Phi_n (x ; t |q) =
\sum_{\stackrel{a+b=n}{a,b \in {\bf Z}_+}}
\frac{(t;q)_a (t;q)_b}{(q;q)_a (q;q)_b} x^{a-b}
\eqno(2.14b)
$$
with the $q$-shifted factorial $(t; q)_a$
defined by Eq.~(2.10a)

Finally, the connection between the $ \Phi_n$'s and the RAI polynomials
is
$$
\Phi_n (e^{i \theta}; t|q) = C_n (\cos \theta; t | q)~.
\eqno(2.14c)
$$
For future reference we give here the
expression [31] for the RAI polynomials in terms of the
$q$-hypergeometric function $_3\phi_2$.
$$
C_n (\cos \theta; t |q) = t^{-n}e^{-in \theta}
\frac{(t^2; q)_n}{(q; q)_n} {_3\phi_2} (q^{-n}, t, te^{2i \theta}; t^2 , 0 |
q, q)
\eqno(2.14d)
$$

The Macdonald scalar product (2.13) now
reduces to the usual scalar product on RAI polynomials
$$
\eqalign{
& <P_m, P_n >_{t,q} = \int_{-1}^{+1} P_m (e^{i \theta} ; t | q) P_n (e^{i
\theta};
t | q ). \cr
& \cdot w (\cos \theta ; t | q ) ( \sin \theta )^{- 1/2} d \cos \theta = \cr
& \delta_{mn}
\frac{(t q^n ; q)_\infty (t q^{n+1}; q)_\infty}{(t^2 q^n ; q)_\infty
(q^{n+1};q)_\infty} \cr }
\eqno(2.15a)
$$
with the weight function $w$ given by
$$
w (\cos \theta ; t |q ) = \frac{1}{2 \pi}
\frac{(e^{2 i \theta} ; q )_\infty (e^{-2i \theta} ; q)_\infty}{(t e^{2i
\theta};q)_\infty (t e^{-2i \theta}; q)_\infty }
\eqno(2.15b)
$$
With the notation (our $l$ is Macdonald's $k$)
$$
t =q^l ~~~{\rm or~~~~~}l = \log t/\log q
\eqno(2.16)
$$
we can also write
$$
\eqalign{
||P_n||^2 & = < P_n, P_n>_{t,q} =
\frac{(q^n t; q)_l}{(q^{n+1}; q)_l} =
\frac{(q^{n+l}; q)_\infty (q^{n+ l + 1}; q)_\infty}{(q^{n+2
l}; q)_\infty (q^{n+1}; q)_\infty } \cr
& = \frac{\Gamma_q (n + 2 l)}{\Gamma_q (n + l)}
\frac{\Gamma_q (n+1 )}{\Gamma_q (n+ 1  + l)} \cr}
\eqno(2.17)
$$
where we used Eq.~(2.10c) and the definition of the $q$-gamma function [28,
29]
$$
\Gamma_q(x) =
\frac{(q; q)_\infty}{(q^x; q)_\infty} (1-q)^{1-x}
\eqno(2.18)
$$
Introducing the Harish-Chandra-like $c-$  function of Macdonald
$$
c(x; l | q) =
\frac{\Gamma_q (x)}{\Gamma_q (x + l)}
\eqno(2.19)
$$
we can finally recast Eq.~(2.12) in the form
$$
||P_n||^2 =
\frac{c(n+ 1; l | q)}{c(n + l; l |q)}
\eqno(2.20)
$$
Through this formula the Macdonald polynomials $P_n$ yield a Macdonald-
Harish-Chandra $c$-function.  We shall make extensive use of this fact in
the following sections.

In Section 4 we shall show that this $c$-function as it emerges from the
Macdonald scalar product via Eq.~(2.20), can be obtained directly from the
large $n$ behavior of the RAI polynomials along the usual lines of
Harish\--Chandra theory [25].

\section{Macdonald's miracles}

In spite of its relative ease, the $A_1$ case considered here, already
exhibits a number of ``miracles", which as shown by Macdonald, generalize
to all admissible pairs of root systems.

A)  We have
$$
\lim_{t=q^{1/2} \rightarrow 1} P_n (e^{i \theta}; t|q) = \left [
\frac{\Gamma (1/2) \Gamma (n+1)}{\Gamma (n + 1/2)} \right ] P_n (\cos \theta
)
\eqno(3.1)
$$
where $P_n  (\cos \theta )$  are the ordinary
Legendre polynomials, the zsf's on
the ordinary compact archimedean symmetric space $SU(2)/SO(2)$, the  2-sphere.

B)  By contrast, for  $q  = 0$ and $t=1/p$ with $p$ a rational prime
$$
P_n (x; {1 \over p} | 0) = \left [ (1 + {1 \over p}
\delta_{n0})^{-1}p^{\frac{n-2}{2}} (1+p) \right ]
\rho_x (n)
\eqno(3.2a)
$$
with
$$
\rho_x(n) =
\frac{x^n(px - x^{-1}) + x^{-n} (x-px^{-1})}{p^{n/2}(p+1)(x-x^{-1})} ,
\eqno(3.2b)
$$
the Mautner-Cartier polynomials [32,33], the zsf's on the non\--compact
$p$-adic
symmetric space
$H^{(p)} = SL(2, {\bf Q}_p )/SL(2, {\bf Z}_p)$,the $p$-adic hyperbolic plane.

{\bf Remark}:   There is a big difference between the interpretations
of the two ``left\--over'' variables $x$ and $n$ in the two cases A) and B)
above.
In the archimedean limit A), the variable $\cos \theta =  (x +
x^{-1})/2$ is the ``radial'' coordinate on the real compact symmetric space
$SU(2)/SO(2)$ and the quantized (angular) momentum variable $n$ is
related to the eigenvalue of the laplacian $n(n+1) (= l ( l +1)$
in more familiar notation).
Things are reversed in the $p$-adic case B).
There the discrete variable $n$ plays the role of ``radial  distance"
on the non-compact $p$-adic symmetric space $H^{(p)}$, which is a
discrete space, a Bruhat-Tits tree [34] (or Bethe lattice).  Conversely,
it
is now the variable $x$ which is related to the eigenvalue of the laplacian
on the tree.
This switch of
variables between the cases A) and B) has a counterpart for all other
root systems [1,2].
Specifically for $q = 0, t = 1/p$ (case B) one obtains
the zsf's of the $p$-adic group G relative to a maximal compact subgroup K
such that the restricted root system of this $(G/K)_{p-adic}$ is
$R^\vee$,
the dual
of the root system R which underlies the real symmetric space $(G/K)_{real}$
whose zsf's one reproduces in the archimedean case A) $(t=q^{ 1/2} \rightarrow
1)$.
In the case at hand, $R= R^\vee = A_1$, so that the difference between the
real and $p$-adic cases is reflected only in the exchanged r\^{o}les of the $x$
and $n$ variables.
This is a very important point which will be further developed in the next
Section.

C)	For $t = 1$, the $P_n$'s take the simple $q$-independent form
$$
P_n (e^{\alpha \over 2} ; 1 | q) = e^{n {\alpha \over 2}} + e^{-n {\alpha
\over 2}}
\eqno(3.3)
$$
in other words they reduce to the $m_n$ s of Eq.~(2.6).

D)  For $q = t$ the $P_n$'s are again $q$-independent and this time they
reduce to the  Weyl characters
$$
P_n (e^{\alpha \over 2}; q|q ) = \chi_n
\eqno(3.4)
$$
with the $\chi_n$ given by Eq.~(2.7).

Macdonald assumed both $q$ and $t$ real.
If we relax this
restriction, we can consider the case of $q$ an $s^{th}$ root of unity.

E)  For $q^s = 1$, the Macdonald polynomials $P_n(y;t |s)$ are
quasi-periodic in $n$:
$$
P_{ns+k} = P_{ns} P_k ~,~~~
n \in {\bf Z}_+ ~, ~k \in \{ 0,1,2, \ldots , s-1 \}
\eqno(3.5)
$$
To see this, recall the recursion relation for Macdonald polynomials
(see [31] Eq.~(2.15))
$$
P_{n+1} = (y+y^{-1})P_n - C_{n-1} P_{n-1}
\eqno(3.6a)
$$
with
$$
C_{n-1} =
\frac{1-t^2 q^{n-1}}{1-t q^{n-1}} ~~
\frac{1-q^n}{1-tq^n} =
\frac{||P_n ||^2}{|| P_{n-1} ||^2}
\eqno(3.6b)
$$
Notice that $q^s =1$ then implies
$$
\begin{array}{c}
C_{ns+k}= C_k \\ C_{ns-1}=0
\end{array}
 ~~~~~~~ n \in {\bf Z}_+, ~
k \in \{0,1,2, \ldots , s-1 \}
\eqno(3.7)
$$
Now from Eqs.~(2.14) $P_0=1, P_1=y+y^{-1}$, so that from Eqs.~(3.6),
(3.7)
it follows that
$$
\eqalign{
P_{ns} & = P_{ns} P_0 \cr
P_{ns+1}& = P_{ns} P_1 \cr }
\eqno(3.8)
$$
Inserting (3.8) and (3.7) into the lin\-ear Eqs.~(3.6a) then yields
the quasi\--perio\-dicity (3.5).

F)  For $t = q^{1/2} \neq 1$, the $P_n$'s become essentially continuous
$q$\--Legendre polynomials, which can be interpreted [15] as
``quasi\--spherical'' functions of the one\--parameter quantum group
$SU(2)_q$.

G)  In the limit $t = q^{(m- 2) /2} \air 1$ the RAI polynomials yield [35]
the
Gegenbauer polynomials, zsf's on the $(m-1)$\--sphere
$S^{m-1} = SO(m)/SO(m-1)$.

\section{Further zonal spherical function-like properties of the
Rogers\--Askey\--Ismail polynomials}

In the remark following properties A) and B) in Section 3, we described the
remarkable interchange of (radial) coordinate and momentum variables  between
the archimedean case A) and the $p$\--adic case B).
This raises the question of what the coordinate and momentum variables
are for
generic values of the two parameters $q$ and $t$.

This question can be answered by noting a remarkable ``self\--duality''
property of the RAI polynomials.
To explain this, let us first observe that the Macdonald polynomials $P_n
(x; t |q)$ yielded familiar spherical functions in the two limiting cases A)
and B) above, only up to the numerical factors in square brackets in
formulae (3.1) and (3.2a).
These inconvenient factors can be eliminated at the expense of relaxing
condition b of Section 2 to rationality in
the variables $t^{1/2}$ and $q$ rather than in
$t$ and $q$.
Then, instead of the $P_n$'s we can define
$$
\Psi_n (e^{i \theta}) =
\frac{\Phi_n (e^{i \theta}; t|q)}{\Phi_n (t^{1/2};t|q)}
\eqno(4.1)
$$
These $\Psi_n$'s, rather than the Macdonald polynomials $P_n$ themselves,
are the candidates for spherical functions of some, as yet hypothetical,
quantum symmetric space.
It is worth noting that
$$
\Phi_n (t^{1/2}; t | q) = t^{-n/2}
\frac{(t^2; q)_n}{(q;q)_n} ~~.
\eqno(4.2)
$$
Define
$$
\nu = \theta / \log q ~~.
\eqno(4.3)
$$
Recalling the definition of $l$ from Eq.~(2.16), and combining it with
Eqs.~(4.1), (4.2), (4.3) and (2.14d), we then obtain
$$
\Psi_n (q^{i \nu }) = q^{- n (2 i \nu + l )/2} {_3 \phi_2} (q^{-n}, q^l , q^{2i
\nu +l}; q^{2l}, 0|q, q) .
\eqno(4.4)
$$
Since the prefactor and the $q$\--hypergeometric [28,29] function $_3
\phi_2$ are
{\em both} invariant under the exchange
$$
- n \leftrightarrow 2 i \nu + l
\eqno(4.5)
$$
it then follows that
$$
\Psi_n (q^{i \nu} ) = \Psi_{-2i \nu - l} (q^{- (n+l)/2})
\eqno(4.6)
$$
where the right hand side is to be understood as obtained by analytic
continuation.
In Eq.~(4.6) the left\--hand side is relevant for the compact case, the
right\--hand side (an analytic continuation) applies to the non\--compact
case.
In particular, this explains the r\^{o}le exchange of the $x$ and $n$ variables
between the two extreme cases A) and B) above ($SU(2)/SO(2)$ is compact,
$SL(2,{\bf Q}_p )/SL(2,{\bf Z}_p)$ is {\em not}).

We can now use Eq.~(4.6) to give a conceptual definition of the
Mac\-don\-ald\--Harish\--Chandra $c$\--function $c(x; l |q)$ of Eq.~(2.19).
Going to large ``distance'' in the non\--compact case, means $n \rightarrow
\infty$ in $\Psi_{-2i \nu - l}(q^{-(n+ l)/2}$).
According to Eq.~(4.6) this means going to large $n$ in $\Psi_n
(e^{i \nu \log q})$.
Using Eqs.~(4.1) and (2.14c), this means going to large $n$ in the RAI
polynomials $C_n (\cos \theta; t|q)$, where $\theta = \nu \log q$.
But this large $n$\--asymptotics of the RAI polynomials follows from the
$q$\--integral representation of these polynomials.
Specifically, for large $n$ [35, 28]
$$
\eqalign{
C_n (\cos \theta; t|q) & = (1-q)^{-l}
\frac{(t;q)_\infty}{(q; q)_\infty} \left [
\frac{\Gamma_q (iu (\theta ))}{\Gamma_q (iu (\theta ) + l )}
e^{-in \theta}  \right . + \cr
& +
\left . \frac{\Gamma_q (-iu (\theta ))}{\Gamma_q (-iu (\theta ) + l )}
e^{in \theta}  \right ] \cr }
\eqno(4.7a)
$$
with
$$
l = \log t / \log q ~~~~~~ {\rm and} ~~~~~u (\theta ) =
\frac{2 \theta}{\log q} ~~.
\eqno(4.7b)
$$
According to Harish\--Chandra we expect the coefficients of $e^{\mp i n
\theta}$ to be $c ( \pm iu ; t | q)$.
Comparing with Eq.~(2.19), we see this is indeed the case.

We are concerned in this paper with interpreting the
RAI polynomials, or more precisely
the $\Psi_n$'s (Eq.~(4.1)) as
zonal spherical functions (zsf's) of a
quantum symmetric space.
In the classical case, a complex valued function $\phi (g), g \in G$ on
a Lie group $G$ is a zsf of $G$ relative to its compact subgroup $K$ if

i) $\phi$ is regular at the identity element $e$ of $G$ and suitably
normalized there $\phi (e) = 1$;

ii) $\phi$ is $K$ biinvariant, i.e., $\phi (k_1 g k_2 ) = \phi (g)$ for all
$g \in G$ and all $k_1, k_2 \in K$;

iii) $\phi$ obeys the functional equation
$$
\phi (g_1) \phi (g_2) = \int_K \phi (g_1 k g_2) d_{Haar} k .
\eqno(4.8)
$$
According to a classical theorem, condition iii) is tantamount to requiring
that $\phi (g)$ be a pull\--back to $G$ of a function on the symmetric space
$G/K$ which is an eigenfunction of each $G$\--invariant differential
operator on $G/K$.
As an example for Legendre polynomials $P_n (\cos \theta )$, the zsf's of
$SO(3)/SO(2)$, Eq.~(4.8) becomes
$$
P_n (\cos \alpha ) P_n (\cos \beta ) = \frac{1}{2 \pi} \int_0^{2 \pi} P_n
(\cos \alpha \cos \beta - \sin \alpha \sin \beta \cos \gamma ) d \gamma
\eqno(4.9)
$$
Now if the RAI polynomials are zsf's of a quantum symmetric space, then we
expect them to obey a relation of the type (4.8).
As a matter of fact {\em they do} [35].

\section{Macdonald polynomials, Sklyanin algebras and $Z_n$-Baxter models}

Our aim is to find the two\--parameter quantum group whose zonal spherical
functions are the Macdonald polynomials for the root system $A_1$ i.e. the
RAI polynomials.
To explain our way of dealing with this question, let us consider, by
analogy, a more familiar problem.
Suppose we are given the Mautner-Cartier polynomials Eq.~(3.26) and we want
to find out whether they are the zsf's of $SL(2, {\bf Q}_p)$ relative to
$SL(2, {\bf Z}_p)$.
The simplest way to do this would be to consider ``$S$\--wave'' scattering
on the $p$\--adic hyperbolic plane
$SL(2, {\bf Q}_p)/SL(2, {\bf Z}_p)$ and to find the corresponding scattering
matrix element
$S_p(u)$, which is expressed in terms of the Jost function $J_p(iu)$
$$
S_p(u) =
\frac{J_p(iu)}{J_p(-iu)}
\eqno(5.1)
$$
If the Mautner-Cartier polynomials {\em are} the appropriate zsf's, then the
Jost function $J_p(iu)$ must coincide with the Harish-Chandra $c$\--function
derived from the large $u$ behavior of the Mautner-Cartier polynomials (for
the chosen value of $p$).
Similar considerations apply to the continuation to complex $n$ of the
Legendre polynomials $P_n(\cos \theta )$.

In our problem we want to see whether the RAI polynomials are spherical
functions of a Sklyanin type quantum group.
To this end we choose a physical system for which the underlying algebra is
of the Sklyanin type.
Then for this system we set up an appropriate scattering problem (of certain
excitations) such that the corresponding Jost function coincides with the
Macdonald\--Harish\--Chandra $c$\--function (Eq.~(2.19)) derived from the RAI
polynomials (see Eqs.~(4.7)).

The appropriate physical system is the $Z_n$\--Baxter model (${\cal B}_n$ for
short) of statistical mechanics on a square lattice [16-19].
The $n^2 \times n^2 ~ R$\--matrix of this model was parametrized by
Belavin [18]
in terms of Jacobi theta functions.
The algebra
which allows a solution of the (intertwining) Yang\--Baxter equations, thus
leading to the existence of infinitely many commuting transfer matrices and
therefore to the integrability of the model,
was studied by Sklyanin [20,21], Cherednik [22,23], and by Odeskii
and Feigin [24], in whose notation the algebra is $Q_{n^2, n-1} ({\cal E}
, \gamma
)$, which we shall call $Q_n$ for short.
Its data are the integer $n$, an elliptic curve ${\cal E}$ and a point
$\gamma$ on ${\cal E}$.
In particular $Q_2 ( = Q_{4,1} ({\cal E}, \gamma ))$ is the original Sklyanin
algebra [20,21] of the 8\--vertex model [16,17].
We do not need the detailed form of the Belavin $R$\--matrix elements.
The essential thing is that the statistical
weights depend on three
independent parameters:  the spectral parameter $z$, the anisotropy parameter
$\gamma$ and the modular parameter $\tau$.
As is customary, we treat $z$ as a variable and $\gamma, \tau$ as
parameters.
In fact it is convenient to treat $n$ as a parameter on equal
footing with $\gamma$ and $\tau$.
Along with ${\cal B}_n$ we also find it useful to think in terms of the
corresponding
$(1+1)$\--dimensional field theoretical models ${\cal M}_n$.
The hamiltonian of ${\cal M}_n$ is obtained from the transfer matrix $T(z)$
of ${\cal B}_n$  through logarithmic differentiation at a special point.
${\cal M}_n$ is known as the generalized magnetic model [36].
Note that ${\cal B}_2$ is Baxter's famous eight\--vertex model, and ${\cal
M}_2$ the familiar XYZ chain.
We shall be interested in the antiferromagnetic regime with finite gap, so
that the ground state is constructed by filling the false (ferromagnetic)
vacuum with quasiparticles.
The partition function $t(z)$ of the ${\cal B}_n$ model in the thermodynamic
limit (the Perron\--Frobenius dominant eigenvalue of $T(z)$) was obtained by
Richey and Tracy [19].
Up to some irrelevant factors, it is of the form
$$
\begin{array}{c}
t(z) = \vartheta  \left [
\stackrel{\scriptstyle{1/2}}{\scriptstyle{1/2}}
 \right ]
 (z - \frac{i \gamma}{\pi} , \tau ) \exp
\left [ -i (\frac{n-1}{n} ) 2 \pi z -  \right . \\
\left .   - i F (z ; \gamma ; n ; \frac{-i \pi \tau}{\gamma} ) \right ] ~~.
\end{array}
\eqno(5.2a)
$$
where
$$
F(z; \gamma ; n , \frac{-i \pi \tau }{\gamma} ) = 2
 \sum_{k=1}^{\infty} {1 \over k}
\frac{{\rm sinh} ~ k \gamma (\frac{-i \pi \tau}{\gamma} - 1 ) }{{\rm sinh}
 ~k \gamma (\frac{-i \pi \tau}{\gamma})}~~
\frac{{\rm sinh} ~k \gamma (n-1)}{{\rm sinh} ~ k \gamma n } \cdot \sin 2 \pi
k z
\eqno(5.2b)
$$
and $\vartheta  \left [
\stackrel{\scriptstyle{1/2}}{ \scriptstyle{1/2}}
\right ] $
is the standard odd theta function with modular parameter
$\tau ( \tau \in i {\bf R}_+ )$.
Notice the remarkable {\em symmetry} of $F(z; \gamma ; n , \frac{- i \pi
\tau}{\gamma })$ in its {\em last} two arguments
$$
F(z; \gamma ; n , \frac{-i \pi \tau}{\gamma}) = F (z; \gamma ; \frac{-i \pi
\tau}{\gamma}, n )~.
\eqno(5.3)
$$
This is our first signal to pay special attention to the variable
$ \frac{-i \pi \tau}{\gamma}$ or its inverse $i \gamma / \pi \tau$.
In fact, it will turn out that precisely this combination
$i \gamma /\pi \tau$ is to be identified with the parameter $l$ in
Macdonald's polynomials (Eq.~(2.16)).
In the next Section
we shall make good use of the symmetry property (5.3).

A remarkable fact [37] in quantum integrable models is that the partition
function, as a function of the spectral variable $z$, coincides up to some
simple factors and/or redefinitions of parameters with a two\--particle
dressed $S$\--matrix, the spectral parameter acquiring the interpretation of
relative rapidity of the scattering particles.
We have to be more specific, there being $n-1$ types of dressed excitations
in ${\cal M}_n$.
We therefore briefly recall the picture of these excitations in the context
of the nested Bethe ansatz (BA).
The ground state, as was already mentioned is found by filling the false
vacuum with $n-1$ types of quasiparticles, each type at its own ``level''.
The momenta are distributed continuously in segments $ [-\pi , + \pi ]$ at
each level.
Excitations are viewed as ``holes'' in these distributions.
The type of physical excitation is determined by the level at which the hole
was created.
In terms of the system of interacting particles on the lattice associated to
the ${\cal M}_n$ model in the usual way, the first level corresponds to
charge excitations, while the others to ``isotopic'' excitations.
The levels are naturally ordered according to the sequence of the nested BA.
It turns out that $t(z)$ of Eq.~(5.2) is essentially
the (scalar)
$S$\--matrix $S_1^{(n)}$ for the scattering of two {\em first} level dressed
excitations.
More precisely
$$
S_1^{(n)} = \exp \left [
-i ( \frac{n-1}{n}) 2 \pi z - i F(z; \gamma; n , \frac{- i \pi \tau}{\gamma})
\right ]
\eqno(5.4)
$$
The $S$\--matrix elements for the scattering of other types of excitations
are more complicated.
Therefore we restrict ourselves to the first level sector and its
$S$\--matrix $S_1^{(n)}$.
At this point it pays to
introduce new variables
$$
\begin{array}{c}
l = \frac{i \gamma}{\pi \tau} \\
q = e^{i 2 \pi \tau} \\
u = - \frac{i z}{\tau} \\
\end{array}
\eqno(5.5)
$$
(notice that $l$ is the combination signaled already in the context of the
symmetry property (5.3);
our $q$ is the usual one, i.e. the square of the one in [19]).
Then we can write $e^{-iF}$ in the form
$$
\exp \left [
-iF(z; \gamma; n , \frac{- i \pi \tau}{\gamma}) \right ] =
\frac{\sigma (i u ; l ; n | q )}{\sigma (-i u ; l ; n | q )}
\eqno(5.6a)
$$
with
$$
\sigma (i u ; l ; n | q ) = \prod_{k=0}^{\infty}
\frac{\Gamma_q (iu + nl (k+1))}{\Gamma_q (iu + nlk + l )}
\frac{\Gamma_q (iu + n l k + 1)}{\Gamma_q (iu + nlk + (n-1)l+1)}
\eqno(5.6b)
$$
Now let $n$ go to infinity.
Using the definition (2.18) and keeping in mind that
as $x \air \infty$ for $q < 1$, we have
$(q^x, q)_\infty \air 1$, it is then readily seen that
$$
\sigma (iu ; l , \infty | q ) = [i u ]_q c (iu; l | q)
\eqno(5.7a)
$$
with $$[iu]_q =
\frac{1 - q^{iu}}{1-q}
\eqno(5.7b)
$$
the ``$q$-analogue'' of $iu$ and $c(iu; l | q)$ the
Macdonald\--Harish\--Chandra $c$\--function for root system $A_1$
Eq.~(2.19)!
Combining Eqs.~(5.7), (5.6a) and (5.4) we find
$$
S_1^{(n)} \left |_{\stackrel{~~~}{n= \infty}} = -
\frac{c(iu; l | q)}{c(-i u; l | q)} \right .
\eqno(5.8)
$$
and we see the Macdonald $c$\--function assuming the r\^{o}le of Jost
function in this scattering process.
This clearly establishes the connection between the $n \air \infty$ limit of
the Sklyanin\--Cherednik\--Odeskii-Feigin algebras $Q_n$, which underlie the
${\cal B}_n$ models on the one hand, and the Macdonald polynomials for the
root system $A_1$ on the other hand.
As was mentioned above, the data for the $Q_n$ algebra are    an elliptic
curve ${\cal E} = {\bf C /Z + Z} \tau_{\cal E}$,
characterized by the
modular parameter $\tau_{\cal E}$, or equivalently
$q_{\cal E} = \exp (i2 \pi
\tau_{\cal E})$,
and a point $\gamma$ on ${\cal E}$.
The data for the set of $A_1$\--Macdonald\--RAI polynomials are the two
parameters $t$ and $q$.
The connection between the elliptic and Macdonald parameters is then
$$
\begin{array}{c}
q = q_{\cal E} \\ t = e^{-2 \gamma}
\end{array}
\eqno(5.9)
$$
the second equation following directly from Eqs.~(2.16) and (5.5).
{\em This connection between $Q_n$ algebras $(n \air \infty )$ and Macdonald
polynomials is our main result.}

At this point the question arises as to why the $n \air \infty$ limit had to
be taken.
On the face of it, all we should have had to deal with should have been the
elliptic algebra $ Q_2$
and the models which it underlies ${\cal B}_2$ and ${\cal M}_2$.
Going to $Q_n, {\cal B}_n , {\cal M}_n$ and then letting $ n \air \infty$ is
like searching for $SU(2)$ inside $SU(\infty )$.
For ordinary Lie groups this would be a detour, for elliptic quantum algebras
this may be needed on account of the complicated coproduct situation [24].
But once in $Q_\infty$, how is it we only found the Macdonald polynomials for
root system $A_1$ and not those for higher $A_n$ root systems?
The point is that  we only looked at the scattering of
{\em two} first level excitations.

After this discussion, we would like to see what would happen, were we
to choose $n=2$, as naively indicated for root system $A_1$, instead of
letting $n \air \infty$.

 From Eqs.~(5.4) - (5.6) we can see that for $n=2$ we find
$$
S_1^{(n)} \left |_{n=2} = q^{iu/2}  \prod_{k=0}^\infty
\frac{c(iu_k - l+1; l | q)}{c(iu_k; l|q)}~ \cdot ~
\frac{c(-iu_k ; l | q)}{c(-iu_k - l+1 ; l|q)}  \right .
\eqno(5.10a)
$$
with
$$
iu_k = iu + l (2k+1) ~~.
\eqno(5.10b)
$$
We see that the building block of $S_1^{(n)} |_{n=2}$ is again the
Macdonald\--Harish\--Chandra $c$\--function of Eq.~(2.19), but this time in a
pattern not as conceptually simple as that of Eq.~(5.8).
Yet we shall have more to say about this case in the next Section.

Whether or not the $n \air \infty$ limit is taken, it would be nice to have
a derivation of the Macdonald\--Harish\--Chandra $c$\--function of
Eq.~(2.19) directly from symmetric spaces constructed from $Q_n$ quantum
groups,
along the standard lines of Harish\--Chandra theory (see e.g. [25]) and
without any reference to the physics of the $Z_n$\--Baxter models.
Conversely it would be of interest to find the geometric object for which
the  infinite product $\sigma (i u; l; n |q)$ of Eq.~(5.6b) is the
$c$\--function.
To steer all this into more familiar territory, notice that in the limit $q
\air 1$ the $q$\--gamma functions in the infinite product reduce to ordinary
gamma functions and the full infinite product (5.6a) becomes essentially
that which appears [38] in the soliton\--soliton scattering $S$\--matrix
in the
sine\--Gordon model, provided one relates our parameter $nl$ to the
sine\--Gordon coupling constant $\beta$ via
$$
n l = 2 \left ( \frac{8 \pi}{\beta^2} - 1 \right )
\eqno(5.11)
$$
Thus the problem of understanding the ``geometric'' interpretation of the
infinite products has as an important special case soliton\--soliton scattering
in the sine\--Gordon model.
Conversely, we can regard the $S$\--matrix given by Eqs.~(5.4)\--(5.6) as a
``$q$\--deformation'' $\Gamma \air \Gamma_q$ of the sine\--Gordon soliton
scattering matrix of [38].
The Sklyanin algebra $(n=2)$ looks like the {\em further} deformation of
quantum $SL(2)$ by a new parameter.

To conclude, let us mention that the expression of the Perron\--Frobenius
dominant eigenvalue of Baxter's zero\--field 8\--vertex model
$( {\cal B}_2)$ transfer
matrix, has been recast in terms of the $c$\--function (2.19).

\section{Interesting special cases}

With the just-established connection between ${\cal B}_n$
or ${\cal M}_n$ systems and
Macdonald polynomials it becomes interesting to see what happens in the
regime in which the polynomials, ``go'' $p$\--adic, i.e., in case B) of
Section 3.
For the $n \air \infty$ situation of Eq.~(5.8)  this corresponds to
$$
q=0 , ~~~~~~ t = e^{-2\gamma} = 1/p ~~.
\eqno(6.1a)
$$
so that
$$
\gamma = \log \sqrt{p}
\eqno(6.1b)
$$
Eqs.~(5.8), (5.5), (2.19) and (2.18) then yield
$$
S_1^{(n)} \left |_{n= \infty, q=0, t=1/p} =
\frac{pe^{i2 \pi z} - 1}{e^{i 2 \pi z} - p} ~,
\right .
\eqno(6.2)
$$
which coincides with the {\em bare} $S$\--matrix in the XXZ model with the
same value of $\gamma$.
Could this result also be obtained from a model on a Bethe lattice with
$p+1$ edges incident at each vertex?

A direct study of ${\cal M}_\infty$ models using the powerful quantum
inverse scattering method or the Bethe ansatz is highly desirable.

The other interesting case is
$$
t = q^{1/2} ~~,
\eqno(6.3a)
$$
so that
$$
l = 1/2 ~~.
\eqno(6.3b)
$$
This corresponds, according to Section 3F, to the familiar {\em
one}\--parameter quantum group $SU(2)_q$.
In the limit $q \air 1$ it then yields the ordinary Lie group $SU(2)$
(Section 3A).
For the $SU(2)_q$ case we have the direct treatment by one of us [12].
An immediate comparison with the results of [12] is not possible, since
there $n = 2$, whereas for us
$$
n \air \infty ~, ~~~~~~~~ l^{-1} = -
\frac{i \pi \tau}{\gamma} = 2
\eqno(6.4)
$$
It is clear from Eq.~(5.4) that for large $n$, the $S$\--matrix depends only
on the function $F(z; \gamma ; n ; \frac{i \pi \tau}{\gamma} )$.
But, as we saw in Eq.~(5.3) this function is symmetric under the interchange
of its last two arguments.
Therefore instead of the case (6.4) we can deal with the equivalent case of
$$n = 2 , ~~~~~~~
\frac{-i \pi \tau}{\gamma} \air  \infty
\eqno(6.5)
$$
which is then in line with [12], provided one replaces (5.5) by
$$
l = \frac{1}{n} ~, ~~~ q = e^{-2 \gamma n} ~~~ u = \frac{\pi
z}{\gamma n} ~~,
\eqno(6.6)
$$
so that, yet again
$$
t = q^l = e^{-2 \gamma} ~~.
\eqno(6.7)
$$
As in [12], we find the XXZ model in this case.

We can also view the \p and $SU(2)_q$ cases directly on the ${\cal B}_2$ or
8\--vertex model or on the equivalent XYZ model, \`{a} la (5.10).
In terms of Baxter's parameters [16] the \p case corresponds to the
6\--vertex model in the
principal antiferroelectric regime with
$$
\Gamma = 1,~~~~ ~ \Delta = -
\frac{p^{1/2} + p^{- 1/2}}{2},
\eqno(6.8)
$$
In terms of the XXZ chain this corresponds to the antiferromagnetic XXZ
chain
$(\Gamma = 1)$ with asymmetry $\Delta$ given by (6.8) (remember $J_x  : J_y
: J_z = 1:
\Gamma : \Delta $).

Similarly the case $t = q^{1/2}$ of the ordinary one\--parameter quantum
group $SU(2)_q$  corresponds to
$$
\Gamma = 0 ~~~~~~~
\Delta = - \frac{1+k}{2 \sqrt{k}}
\eqno(6.9)
$$
where $k$ is the modulus of the elliptic Jacobi functions of nome $q$.
This is the XZ model.
If we now let $t = q^{1/2} \air 1$, corresponding to the ordinary $SU(2)$
Lie group (Case A of Section 2)
then the comodulus $k' \air 0$, so that the modulus $k \air 1$ and $\Delta
\air - 1$.

\section{Applications and generalizations}

a) A large class of elliptic quantum algebras has already been brought into
play in the context of the $Z_n$-Baxter models and the simplest Macdonald
polynomials.
The question then naturally arises as to a full classification of
elliptic quantum
symmetric spaces, in correspondence with admissible pairs of root
systems.

There is one more aspect to this.  The
parameters $q,t$ of Macdonald translate on the ``Sklyanin side'' into an
elliptic curve and a point on it.  Could one make the connection with
elliptic curves explicit directly on the ``Macdonald side''?

b) For generic $t$ and $q$, the orthogonal RAI polynomials obey of course,
a three term recursion relation (Eqs.~(3.6)).
In the \p regime $(q=0, ~ t = 1/p)$ this recursion relation becomes precisely
the condition that the zsf's be eigenfunctions of the laplacian.
On the Bruhat\--Tits tree, corresponding to this case, the laplacian has a
simple interpretation as a difference operator obeying the mean value
theorem.
It is then natural to expect that
for generic $t$ and $q$,
the recursion relation (3.6)
also corresponds to the requirement that the RAI
polynomials be eigenfunctions of the laplacian on some ``non\--arboreal''
{\em discrete} space, which reduces to a tree in the \p regime.
It would be nice to find a simple geometric description of this generic
discrete space which in the \p case becomes a tree whereas for $t = q^{1/2}
\air 1$ becomes a (continuous) sphere of (real) dimension 2.
In short then it would be interesting to have a direct geometric picture of
the quantum symmetric  space, not only of its zonal spherical functions.

c) The interpolation between real and $p$-adic symmetric spaces by
varying the  parameters $q$ and $t$, makes one recall another real-$p$-adic
connection, at the adelic level [26], via Euler products [38].
In fact, for $q = 0, t = 1/p$ the $c$-function is a ratio of local ($p$-adic)
zeta functions.
$$
\zeta_p (s) = \frac{1}{1 - p^{-s}}
\eqno(7.1)
$$
whereas for $t=q^{1/2} \air 1$, the $c$-function is a ratio of (real) local
zeta-functions
$$
\zeta_\infty (s) = \pi^{-s/2} \Gamma \left ( {s \over 2} \right )
\eqno(7.2)
$$
Taking the Euler product yields the adelic zeta-function
$$
\Lambda (s) = \zeta_\infty (s) \prod_p \zeta_p (s) = \pi^{-s/2} \Gamma ( {s
\over 2} ) \zeta (s)
\eqno(7.3)
$$
which involves the Riemann zeta function $\zeta (s)$
and obeys the simple functional equation $\Lambda (s) = \Lambda (1 - s)$.
Can this construction be
$q$-deformed?
Is there such a thing as a ``$q$-Euler product''?  To answer
these questions, notice that the Euler product runs one of the Macdonald
variables (namely $t$) over all reciprocal prime values, while the other
stays fixed.
As the other Macdonald variable one can choose either $q$
or $l  = -\log p/\log q$, as both of them stay fixed at zero.  It will turn
out that for us the sensible choice is
$l$.
So we view an Euler
product as a product over $t = 1/2, 1/3, 1/5,\ldots$ while $l$ is fixed at
zero.
A {\em deformed} Euler product then should do the same but with $l$
fixed  at some value other than zero.

At this point we have to find the deformations of the local zeta
functions $\zeta_p(s)$ and $\zeta_\infty (s)$.
If we call $\zeta   (s;t,l)$ the(two-parameter) ``deformed local zeta
function'', then we must impose
$$
\zeta (s; {1 \over p} , 0) = \zeta_p (s)
\eqno(7.4)
$$
and
$$
\zeta (s; 1, 1/2) = \zeta_\infty (s)
$$
It is easy to see that $(q = t^{ 1/l })$
$$
\zeta (s; t, l ) = \pi^{q l s} \Gamma_{t^{ 1/l}} ( l s)
\eqno(7.5)
$$
fits the bill.
We have
$$
\eqalign{
& \zeta (s; t, l) = \pi^{-q l s}
\frac{(t^{1/l}; t^{1/l})_\infty (1-t^{1/l})^{1-l s}}{(t^s ,
t^{1/l})_\infty } = \cr
& = \pi^{-q l s} (1 - t^{1/l})^{1-l s}
\frac{\prod_{n=1}^{\infty} (1-t^{n/ l})}{\prod_{n=0}^{\infty}(1-t^{s+ n/l})}
 \cr }
\eqno(7.6)
$$
which for $t = 1/p < 1 , ~ l \air 0+ , ~ q \air 0$ does indeed become
$\frac{1}{1-p^{-s}} = \zeta_p(s)$.
On the other hand, for
$l = 1/2 , ~ t \air 1$ (so that also $q \air 1$),
$$
\zeta (s; 1, 1/2) = \pi^{-s/2} \Gamma \left ( {s \over 2} \right )
= \zeta_\infty (s)
\eqno(7.7)
$$
To get a meaningful $q$-Euler product starting from (7.5), we have to
unload the $\pi^{-q l s}$ factor.
So the $q$-deformed Euler factor will be $\pi^{q l s} \zeta (s;t,l )
|_{l = {\rm fixed}, t = 1/p}$,
and our ``$q$-Euler product'' or more appropriately   ``$l$-Euler product''
will be $(q = t^{1/l})$
$$
\eqalign{
& E(s; l) = \prod_p \pi^{q l s} \zeta (s; {1 \over p}; l ) = \cr
& = \prod_p
\frac{(1-p^{- 1/l})^{1-l s} \prod_{n=1}^{\infty} (1-p^{-n/l}
)}{\prod_{n=0}^{\infty} (1-p^{-s-n/l} )} =
\frac{\zeta (l^{-1})^{1-l s} \prod_{n=1}^{\infty} \zeta (n
l^{-1})}{\prod_{n=0}^{\infty} \zeta (s + n l^{-1})} \cr }
\eqno(7.8)
$$
an interesting combination.

The special r\^{o}le in all this of the $q$\--gamma function $\Gamma_q (ls)$
as ``interpolator'' between the local zeta functions at the finite and
infinite places, can be better understood by tracing it back to a remarkable
property of the $q$\--exponential and to a remarkable property of Jackson's
$q$\--integral.
The point is that the $q$\--gamma function admits a $q$\--integral
supresentation [29], which upon a standard change of variables turns into a
$q^l (=t)$\--integral representation.
This is essentially a $t$\--Mellin transform of $e_q(-x^{1/l})$.
Here the $q$\--exponential $e_q(y)$ is defined as [28], [29]
$$
e_q (y) =
\sum_{a=0}^\infty
\frac{y^a}{[a]_q!}~, ~~
[a]_q =
\frac{1-q^a}{1-q} ~, ~~
[a]_q! =
[a]_q [a-1]_q \cdots [1]_q
\eqno(7.9)
$$
Not surprisingly for $t \air 1 , ~ l \air 1/2$ (the archimedean regime) this
$t$\--Mellin transform reduces to the ordinary real Mellin transform and
$e_q (-x^{1/l})$ becomes the real gaussian (a well known representation of
the gamma function of half\--argument).
The \p regime, surprisingly allows a similar interpretation.
A $t$\--integral is a sum [29]
$$
\int f(x)d_t x =
\sum_{n=- \infty}^{+ \infty} f(t^n) [(1-t)t^n] ~.
\eqno(7.10)
$$
In the \p regime $q \air 0 ~ t \air 1/p$, the factor in square brackets
coincides with the volume of the ``shell''
$I_n = \left \{ \xi \in {\bf Q}_p , ~
\ze = p^{-n} \right \}$
of \p integration, so that the sum over $n$, itself amounts to an
integration over ${\bf Q}_p$ of the complex function $f ( \ze )$ of the \p
variable $\xi$ ($f$ depends on $\xi$ only through its norm $\ze$).
Under the sum (7.10), $e_q (-x^{1/l})$ becomes $e_q (-t^{n/l}) = e_q (-q^n)$
(since $t = q^l$).
As $q \air 0$, the $q$\--analogues of all nonnegative integers go to one, so
that $e_q (-q^n)$ behaves like the geometric sum
$\sum_{a=0}^\infty (-q^n)^a = (1 + q^n)^{-1}$
and thus equals one for $n > 0$ and zero for $n$ negative.
When the case $n = 0$ is included, this shows that
the function
$~ e_q (-x^{1/l})$,
under $t$\--integration $d_tx$,
translates into the characteristic function $\chi_p
(\xi )$ of the \p integers
$$
\chi_p (\xi ) =
\left \{
\begin{array}{c c c}
1 & {\rm for} & \ze \geq 1 \\
0 & {\rm for} & \ze > 1
\end{array}
\right .
\eqno(7.11)
$$
under \p integration.
But this $\chi_p (\xi )$ is the ``\p gaussian'', that is the Fourier self
similar complex\--valued function of  the \p  variable $\xi$.

We thus come to realize that $t$\--integration ``interpolates'' between real
Riemann integration and \p integration, while at the same time
the function $e_q (-x^{1/l})$ plays
the r\^{o}le of a ``quantum gaussian'' which interpolates between the
ordinary real gaussian and the step\--functions $\chi_p (\xi )$, the \p
gaussians.
All this clearly begs for a $q$\-- and/or $l$\--deformation of Tate's
Fourier analysis on local fields.

d)  Does any of this work bear on string theory?  Yes, we can
construct {\em two-parameter} deformations of string theory which for
$t=1/p,~ q=0$ reproduce the known $p$-adic strings, for $t=q^{1/2} \air 1$,
the ordinary Veneziano string,  and for $t=q^{1/2} \neq 1$ involve
$q$\--strings.
We shall return to this elsewhere.

\medskip

We wish to thank S.~Bloch, L.~Chekhov, L.~Mezincescu and M.~Olshanetsky for
valuable discussions.

\medskip

\centerline{REFERENCES}

\begin{enumerate}

%1
\item Macdonald, I.~G.: in {\em Orthogonal Polynomials:
Theory and Practice}, P.~Nevai ed., Kluwer Academic Publ., Dordrecht, 1990

%2
\item Macdonald, I.~G.: Queen Mary College preprint 1989

%3
\item Freund, P.~G.O.: in {\em Superstrings and Particle Theory},
L.~Clavelli and B.~Harms eds., World Scientific, Singapore, 1990

%4
\item Freund, P.~G.~O.:
Phys. Lett. {\bf 257B}, 119 (1991)

%5
\item Freund, P.~G.~O.:
in {\em Quarks, Symmetries and Strings, A Symposium in Honor of Bunji
Sakita's 60th Birthday}, M.~Kaku, A.~Jevicki and K.~Kikkawa eds., World
Scientific, Singapore, 1991

%6

\item Brekke, L., Freund, P.~G.~O., Olson, M., and Witten, E.: Nucl. Phys. {\bf
B302}, 365 (1988)

%7
\item Zabrodin, A.~V.: Comm. Math. Phys. {\bf 123}, 463 (1989)

%8
\item Coon, D.~D.: Phys. Lett. {\bf 29B}, 1422 (1969)

%9
\item Romans, L.~J.: preprint USC-88/HEP014 (1988)

%10
\item Olshanetsky, M.~A., and Perelomov, A.~M.: Phys. Reports
{\bf 94}, 313 (1983)

%11.
\item Wehrhahn, R.~F.: Phys. Rev. Lett. {\bf 65}, 1294 (1990)

%12
\item Zabrodin, A.~V.: Moscow preprint 1991

%13
\item Ueno, K.: Proc. Jap. Acad. {\bf 66A}, 42 (1990)

%14
\item Vaksman, L.~L., and Korogodsky, L.~I.: Moscow preprint (1990)

%15
\item Koornwider, T.: in {\em Orthogonal Polynomials Theory and Practice},
P.~Nevai, ed., Kluwer Academic Publ., Dordrecht, 1990

%16
\item Baxter, R.~J.: {\em Exactly Solved models in Statistical
Mechanics},
Acad. Press, N.Y., 1982

%17
\item Gaudin, M.: {\em La Fonction D'Onde de Bethe}, Masson, Paris 1983

%18
\item Belavin, A.~A.: Nucl. Phys. {\bf B180}, 189 (1981)

%19
\item Richey, M.~P., and Tracy, C.~A.: J. Stat. Phys. {\bf 42}, 311 (1986)

%20
\item Sklyanin, E.~K.: Funk. Anal. Appl. {\bf 16},263 (1982)

%21
\item Sklyanin, E.~K.: Funk. Anal. Appl. {\bf 17}, 273 (1983)

%22
\item Cherednik, I.~V.: Yad. Fiz. {\bf 36}, 549 (1982)

%23
\item Cherednik, I.~V.:
Funk. Anal. Appl. {\bf 19}, 77 (1985)

%24
\item Odeskii, A.~V., and Feigin, B.~L.: Funk. Anal. Appl. {\bf 23}, 207
(1989)

%25
\item Helgason, S.: {\em Topics in Harmonic Analysis on Homogenous Spaces},
Birkh\-\"{a}user, Basel, 1981

%26.
\item Weil, A.: {\em Adeles and Algebraic Groups},
Birkh\"{a}user, Basel, 1982

%27
\item Humphreys, J.~E.: {\em Introduction to Lie Algebras and Representation
Theory}, Springer, Berlin, 1970

%28
\item Gasper, G., and Rahman, M.: {\em Basic Hypergeometric Series},
Cambridge
Univ. Press, Cambridge, 1990

%29
\item Exton, H.: {\em q-Hypergeometric Functions and Applications}, John
Wiley, N.Y., 1983

%30
\item Rogers, L.~J.: Proc. London Math. Soc. {\bf 26}, 318 (1895)

%31
\item Askey, R., and Ismail, R.~E.~H.: in {\em Studies in Pure Math.}
P.~Erd\"{o}s, ed.
Birkh\"{a}user, Basel, 1983

%32
\item Mautner, F.: Am. J. Math. {\bf 80}, 441 (1958)

%33
\item Cartier, P.: {\em Proc. Symp. Pure Math.} Vol. {\bf 26},
A.~M.~S.~Providence, 1973

%34
\item Bruhat, F., and Tits, J.: Publ. Math. I.H.E.S. {\bf 41}, 5 (1972)

%35
\item Rahman, M., and Verma, A.: SIAM J. Math. Anal. {\bf 17}, 1461 (1986)

%36
\item Kulish, P.~P., and Reshetikhin, N.~Yu.: Soviet Phys. JETP {\bf
53}, 108 (1981)

%37
\item Zamolodchikov, A.~B.:
Comm. Math. Phys. {\bf 69}, 165 (1979)

%38
\item Zamolodchikov, A.~B., and
Zamolodchikov, Al.~B.: Ann. Phys. (N.Y.) {\bf 120}, 253 (1979)

%39
\item Langlands, R.~P.: {\em Euler Products}, Yale University Press, New
Haven, 1971

%40
\item Tate, J.: Thesis, Princeton 1950, reprinted in {\em Algebraic Number
Theory}, J.~W.~S.~Cassels and A.~ Fr\"{o}hlich, eds., Academic Press, N.Y.,
1967

%41
\item Gel'fand, I.~M., Graev, M.~I., and Pyatetskii\--Shapiro, I.~I.:
{\em
Representation Theory and Automorphic Functions}, Saunders, London, 1966

\end{enumerate}

\end{document}